\documentclass[prd,amsmath,amssymb,longbibliography,superscriptaddress,twocolumn,nofootinbib,10pt]{revtex4}

\pdfoutput=1

\usepackage{graphicx}
\usepackage{dcolumn}
\usepackage{bm}
\usepackage{amssymb}
\usepackage{latexsym}
\usepackage{booktabs}
\usepackage{amsmath}
\usepackage{multirow}
\usepackage[colorlinks=true, linkcolor=blue, citecolor=blue]{hyperref}

\newcommand{\Mpc}{\mathrm{~km~s^{-1}~Mpc^{-1}}}
\begin{document}

\title{Simultaneous measurements on cosmic curvature and opacity using latest HII regions and $H(z)$ observations}

\author{Ying Yang}
\affiliation{Hunan Provincial Key Laboratory of Intelligent Sensors and Advanced Sensor Materials, School of Physics and Electronics, Hunan University of Science and Technology, Xiangtan 411201, China}
\author{Tong-hua Liu}
\email{liutongh@yangtzeu.edu.cn}
\affiliation{School of Physics and Optoelectronic, Yangtze University, Jingzhou 434023, China}
\author{Jiayuan Huang}
\affiliation{School of Physics and Optoelectronic, Yangtze University, Jingzhou 434023, China}

\author{Xiaolan Cheng}
\affiliation{School of Physics and Optoelectronic, Yangtze University, Jingzhou 434023, China}

\author{Marek Biesiada}
\affiliation{National Centre for Nuclear Research, Pasteura 7, PL-02-093 Warsaw, Poland}
\author{Shu-min Wu}
\email{smwu@lnnu.edu.cn}
\affiliation{Department of Physics, Liaoning Normal University, Dalian 116029, China}

\begin{abstract}
The different spatial curvatures of the universe affect the measurement of cosmological distances, which may also contribute to explaining the observed  dimming of type Ia supernovae. This phenomenon may be caused by the opacity of the universe. Similarly, the opacity of the universe can lead to a bias in our measurements of curvature.  Thus, it is necessary to measure cosmic curvature and opacity simultaneously. In this paper, we propose a new model-independent method to simultaneously measure the cosmic curvature and opacity by using the latest observations of HII galaxies acting as standard candles and the latest Hubble parameter observations. The machine learning method-Artificial Neural Network is adopted to reconstruct observed Hubble parameter $H(z)$ observations. Our results support a slightly opaque and flat universe at $1\sigma$ confidence level by using previous 156 HII regions sample. However, the negative curvature is obtained by using the latest 181 HII regions sample in the redshift range $z\sim 2.5$.  More importantly, we obtain the simultaneous measurements with precision on the cosmic opacity $\rm\Delta\tau\sim 10^{-2}$ and curvature $\rm\Delta\Omega_K\sim 10^{-1}$.  A strong degeneracy between the cosmic opacity and curvature parameters is also revealed in this analysis.

\end{abstract}

\maketitle

\section{Introduction}

Modern cosmology accepts that our universe is undergoing accelerated expansion in its current phase, a conclusion supported by the most direct evidence yet that Type Ia supernovae (SN Ia) have been unexpectedly observed fainter than expected in a slowing universe
\cite{Riess98,Perlmutter99,Scolnic18}. Several other different theories or mechanisms have been proposed to explain the observed type Ia supernova dimming \cite{Qi17,Xu18}, in addition to a new cosmological component that applies negative pressure (the cosmological constant is the simplest candidate) \cite{Ratra88,Caldwell98,Cao11a,Cao13a,Cao14,Ma17,Qi18}. Shortly after the discovery of cosmic acceleration, it was proposed that the observation of SN Ia might be affected by the non-conservation of the number of photons in the emitted beam. Such so-called cosmic opacity may be due to many possible
mechanisms. The standard mechanism is usually explained as the photons moving through the Milky Way, intervening
galaxies, and the host galaxy being absorbed or scattered by dust particles \cite{Tolman30}. Some other exotic
mechanisms for cosmic opacity discuss the conversion of photons into
gravitons \cite{Chen95}, light axions in the presence of
extragalactic magnetic fields \cite{Csaki02,Avgoustidis10,Jaeckel10}, or Kaluza-Klein modes
associated with extra-dimensions \cite{Deffayet00}.  Various astrophysical mechanisms, such as gravitational lensing and dust extinction, may cause the non-conservation of the number of photons from the view of observation and thus be interpreted as the opacity of the universe.

Although recent works on the opacity of the universe has shown that the universe appears to be transparent \cite{2018PhRvD..97b3538H,2015PhRvD..92l3539L}, some of the growing crises in cosmology, such as Hubble tension \cite{2020A&A...641A...6P,2019ApJ...876...85R,2017NatAs...1E.169F} and the cosmic curvature \cite{2020NatAs...4..196D,2021APh...13102605D,2021PhRvD.103d1301H}, have emerged.  This suggests us to reanalyze and treat the mechanism of cosmic opacity. For instance, if the universe is opaque, the photon from a distant light source reaches the ground and is observed, and its flux received by the observer will be reduced by a factor of $e^{\tau/2}$ ($\tau$ is optical depth, and $\tau<0$ means an opaque universe). According to the relationship between flux and luminosity distance, when the flux decreases, the luminosity distance increases, which means that the Hubble constant decreases. This provides a new way to solve the growing Hubble tension problem \footnote{There is $4.4\sigma$ tension between the Hubble constant ($H_0$) measurements inferred within lambda cold dark matter ($\rm{\Lambda}$CDM) from cosmic microwave background (CMB) anisotropy (temperature and polarization) data \cite{2020A&A...641A...6P} and the local measurement of Hubble constant by the \textit{Supernova $H_0$ for the Equation of State} collaboration (SH0ES) \cite{2019ApJ...876...85R}. }  (more works on Hubble tension please see the references \cite{2019PhRvL.122v1301P,2019PhRvL.122f1105F,2021ApJ...912..150D,2022A&A...668A..51L} and  references therein).
More importantly, some recent works suggested that the $H_0$ tension could possibly be caused by the inconsistency of spatial curvature between the early-universe and late-universe \cite{2021APh...13102605D,2020NatAs...4..196D,2021PhRvD.103d1301H}. Similarly, the opacity of the universe can lead to a bias in our measurements of curvature. Thus, to better understand the growing crises in cosmology, it is necessary to explore the relation between the curvature and the opacity of our universe, and seek the cosmological model-independent method that constrains cosmic curvature and opacity simultaneously. Furthermore, cosmic opacity might be an important source of systematic errors in this respect, and it becomes increasingly important to quantify the transparency of the universe.

From a theoretical point of view, two luminosity distances are required to determine the cosmic opacity parameters, one is the luminosity distance that is not affected by the cosmic opacity (denotes $D_{\mathrm{L,true}}$), and the other is the luminosity distance that is affected by the cosmic opacity (denotes $D_{\mathrm{L,obs}}$). A opacity parameter $\tau$ can be introduced to describe the optical depth associated with cosmic absorption $D_{\mathrm{L,obs}}= D_{\mathrm{L,true}}\cdot e^{\tau/2}$. The luminosity distance that is not subject to cosmic opacity is usually derived from  measurements of differential ages of passively evolving galaxies the so-called cosmic chronometers (CC) \cite{Holanda13,Liao13,Jesus17}. The comoving distance is obtained by integrating the Hubble parameter, and then the luminosity distance is obtained by using the distance duality relation. Combination of opacity-dependent luminosity distances derived from SN Ia observations \cite{2017ApJ...847...45W} or other sources (such as quasars \cite{LiuT21a}) and opacity-independent luminosity distances inferred from Hubble parameter observations,
one can  directly measure the cosmic opacity. However, it should be stressed that the luminosity distance obtained using the comoving distance implies a strong assumption that the universe is flat.

In this work, we focus on the latest observations of Hubble parameter and HII galaxies (HIIGx) and Giant extragalactic HII regions (GEHR) to both constrain the cosmic curvature and opacity. There are two reasons to focus on such a dataset combination. Firstly, the redshift ranges of the latest HIIGx and HII regions sample and observations of Hubble parameter are roughly consistent, and can reach a relatively high redshift range $z\sim 2.5$. Another reason is that the use of HIIGx and HII regions observations is based on the strong correlation between the luminosity $L(\rm{H\beta})$ in $\rm{H\beta}$ lines
and the ionized gas velocity dispersion $\sigma$. This means that the HIIGx and HII regions observations have the advantage of being sensitive to the nonconservation of photon number. However, it is very difficult considering that both luminosity distances should be measured at the same redshift. Therefore, we will reconstruct the Hubble parameter $H(z)$ measurements using the Artificial Neural Network (ANN) method. The ANN method is a kind of machine learning technique which is good at regression, and  has been widely used in astronomical research in recent years \cite{Peel19,Wang20a,Wang2b,LiuT21b}. Wang \textit{et al.} have used the ANN method to reconstruct Hubble parameter $H(z)$ and  proved its reliability and superiority in characterizing data uncertainties \cite{Wang20a}.

This paper is organized as follows: in Section 2, we present the methodology of measuring cosmic curvature and opacity, and the data
used in this work. In
Section 3, we show our results and discussion. Finally, the main conclusions are
summarized in Section 4.

\section{ Data  and Methodology}\label{sec:data}
The foundation of modern cosmology is based on the basic principles of cosmology, i.e., the universe is homogeneous and isotropic at large scales. The Friedmann-Lematre-Robertson-Walker (FLRW) metric is appropriate to describe this situation in the universe and reads
\begin{equation}
ds^2=dt^2-\frac{a(t)^2}{1-Kr^2}dr^2-a(t)^2r^2d\Omega^2,
\end{equation}
where $a(t)$ is the scale factor, and $K$ is dimensionless curvature taking one of three values $\{-1, 0, 1\}$ corresponding to a close, flat and open universe. The cosmic curvature parameter $\rm\Omega_K$ is related to $K$ and the Hubble constant $H_0$, as $\rm{\Omega_K}$ $=-c^2K/a^2_0H_0^2$.

Under the FRLW metric, the luminosity distance $D_L$ and the comoving distance $D_C$ satisfy the following relation \cite{rule1,rule2}
\begin{equation}\label{eq2}
\frac{D_L}{1+z} = \left\lbrace \begin{array}{lll}
\frac{D_H}{\sqrt{|\rm\Omega_{\rm K}|}}\sinh\left[\sqrt{|\rm\Omega_{\rm K}|}{D_C/D_H}\right]~~{\rm for}~~\rm\Omega_{K}>0,\\
D_C~~~~~~~~~~~~~~~~~~~~~~~~~~~~~~~~~~~{\rm for}~~\rm\Omega_{K}=0, \\
\frac{D_H}{\sqrt{|\rm\Omega_{\rm K}|}}\sin\left[\sqrt{|\rm\Omega_{\rm K}|}D_C/D_H\right]~~~~{\rm for}~~\rm\Omega_{K}<0,\\
\end{array} \right.
\end{equation}
where $D_H=c/H_0$ is Hubble radius, and the comoving distance
\begin{equation}
D_C=c\int^z_0\frac{dz'}{H(z')},
\end{equation}
where $c$ represents the speed of light, and $H(z')$ denotes the Hubble parameter at redshift $z'$.

In the following, we will briefly introduce the methodology of deriving two different luminosity distances, i.e.,
opacity-dependent luminosity distance from the latest observation of HII galaxies and extragalactic HII regions, as well as opacity-independent luminosity distance inferred from the current
$H(z)$ data.
\subsection{Opacity-dependent luminosity distances from HII galaxies and extragalactic HII regions}

To accurately measure distances in the distant universe, we always turn to standardized sources, such as SN Ia acting standard candle \cite{9}. Recently, more distant quasars as standard candles by using the nonlinear relation between their intrinsic UV and the X-ray luminosities have obtained great attention \cite{13,LiuT21a,15}. High redshift objects are always interesting, because they contain important information about the physical processes of the early universe. The HII galaxies and extragalactic HII regions constitute a large fraction of population that can be observed up to very high redshifts \cite{18,19,20}, beyond the feasible limits of supernova studies.

 It is well known that the luminosity
$L($H$\beta)$ in H$\beta$ and the ionized gas velocity dispersion
$\sigma$ of HII galaxies and extragalactic HII regions may have a
quantitative relation (be known as ``$L$--$\sigma$" relation). The
physics behind this relation is based only on a simple idea, i.e.,
as the mass of the starburst component increases, the number of
ionized photons and the turbulent velocity of the gas may both
increase as well. Melnick \textit{et al.} first found that the
scatter of ``$L$--$\sigma$" relation is very small and has the
capability to determine the cosmological distance independent of
redshift\cite{21}. More specifically, based on the measured flux
density (or luminosity) and the turbulent velocity of the gas, one
can infer the luminosity distance directly.  Whereafter, the
validity of the ``$L$--$\sigma$" relation acting as the standard
candle and its possible cosmological applications have been
extensively discussed in the literatures \cite{22,23,24}.

The ``$L$--$\sigma$" relation between the luminosity $L($H$\beta)$
in H$\beta$ of a source and its ionized gas velocity dispersion is quantified as \cite{19}
\begin{equation}
\log L ( \textrm{H}\beta) = \alpha \log \sigma(\textrm{H}\beta)+\kappa,
\end{equation}
where $\alpha$ and $\kappa$ are the slope and intercept, respectively. The
$L(\textrm{H}\beta)$ is the luminosity in $\rm{H\beta}$ lines. One can use a general equation
$L(\textrm{H}\beta) = 4\pi D^{2}_L F(\textrm{H}\beta)$ and observations of the extinction corrected fluxes
 $F(\textrm{H}\beta)$ to obtain it. Thus, the new relation between the observed flux
density and luminosity distance can be written as
\begin{equation}
\log D_{\rm{L},\rm{HII}}(z) = 0.5[\alpha \log \sigma(\textrm{H}\beta)- \log F ( \textrm{H}\beta)+\kappa]-25.04.
\end{equation}

Up to now, the catalog of spectral and astrometric data from the HII galaxies and extragalactic HII regions contains more than 100 sources whose statistical properties can be preliminarily considered in cosmology.  The current observations of 156 HII objects (denoted as 156 HII sample) compiled by Terlevich \textit{et al.} \cite{19}
contain 25 high redshift HII galaxies sources, 107 local HII galaxies sources, and 24 extragalactic HII regions sources covering redshift range $0<z<2.33$. This dataset is larger than the source
samples used by Plionis \textit{et al.} \cite{22} and is more
complete than the high redshift data used by Melnick \textit{et al.}
\cite{21}.  Recently, Gonz{\'a}lez-Mor{\'a}n \textit{et al.} \cite{25} reported new observations of 41
high-z $(1.3\leq z \leq2.6)$ HIIG objects which are based on new VLT-KMOS high spectral resolution observations. The new 41 data for high$-z$ objects in addition to the sample in work \cite{25}, giving a total of 74 high$-z$ HIIG objects. In this work, in addition to the samples with 74 high-z HIIG, the complete samples we used also included 107 data local HII galaxies sources, which is consistent with the work in reference \cite{25}. The full sample contains 181 sources covering the redshift range $0<z<2.55$ (denoted as 181 HII sample). Full information (including redshift, flux density in H$\beta$ line, and turbulent velocity with corresponding uncertainties) about the sample of 181 HII regions can be found in  {Ref.} \cite{25}.

Similar to SN Ia applied to cosmology, the slope $\alpha$ and intercept $\kappa$
parameters should also be optimized with the assumed cosmological
model parameters, but this method is cosmological model dependent \cite{20}. The intercept $\alpha=5.022\pm0.058$ and slope $\kappa=33.268\pm0.083$ of the relation are estimated for the sample of 107 local
 HIIG published in work \cite{26a}
and 36 GEHR described in work \cite{26b} using the extinction curve \cite{26c}. We adopt these values with their
corresponding uncertainties to get the luminosity distance and make sure they are cosmological model independent.  The distance modules of the two samples and the corresponding errors are shown in Fig. 1.
\begin{figure}
\begin{center}
\includegraphics[width=0.9\linewidth]{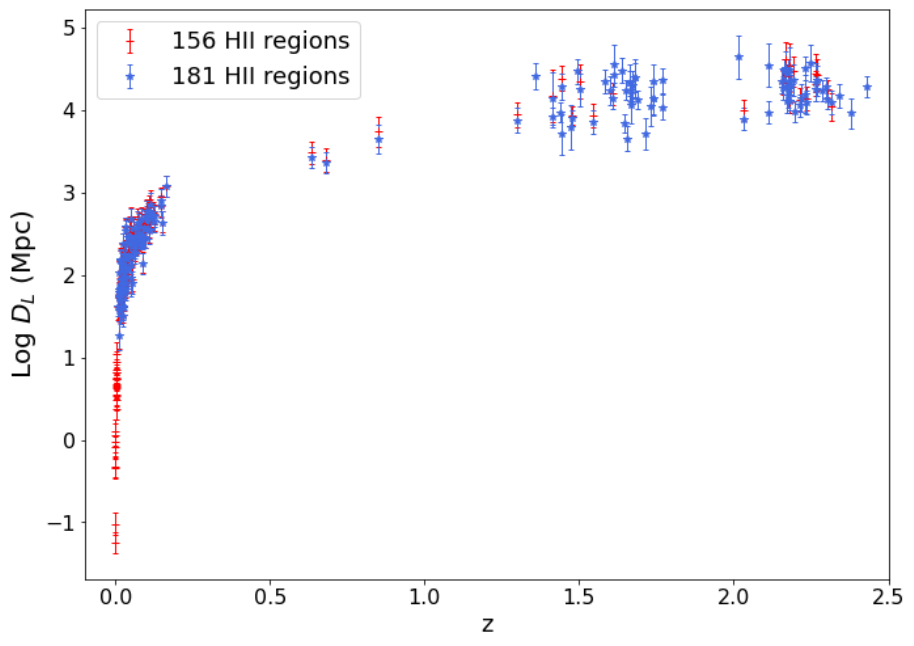}
\end{center}
\caption{ The $\log D_{L,HII}(z)$ with $1\sigma$ errors from HII regions observations. The red and blue dots  represent the 156  and 181 HII  regions observational datasets, respectively. }
\end{figure}
\subsection{Opacity-independent luminosity distances from Hubble parameter measurements}
In order to obtain opacity-independent luminosity distances, we seek the Hubble parameter observations.
In general, the Hubble parameter measurements can be derived by the differential
ages of passively evolving galaxies called {CC}, which is suggested by \cite{Hz2}.  This method is based on the differential relationship between Hubble parameters and redshift
\begin{equation}
H(z)=-\frac{1}{1+z}\frac{dz}{dt},
\end{equation}
where $dz/dt$ is the time derivative of redshift. The CC data is mainly obtained by measuring the different ages of the red-envelope galaxies, a method known as the different ages method. The aging of stars can be seen as an indicator of the aging of the universe. The spectra of stars can be converted into information about their ages because the evolution of stars is well known. Since stars cannot be observed individually on a cosmic scale, spectra of galaxies with relatively uniform stellar populations are often used. This method relies on the detailed shape of the galaxy spectra but not on the galaxy luminosity. We assume cosmic opacity is not strongly wavelength-dependent in the (relatively narrow) optical
band and thus CC data are opacity-independent and cosmological model-independent \cite{2010JCAP...10..024A,Li13,2015PhRvD..92l3539L}.
The full information for total number of 32 CC is listed in Table 1 of the literature \cite{Hz1}. The 32 CC sample covers the redshift range of $0.07\leq z<\leq1.965$ \cite{HzCC1,HzCC2,HzCC3,HzCC4,HzCC5,HzCC6}.

Another method is to detect the radial baryon acoustic oscillation (BAO) features by using galaxy surveys and Ly$-\alpha$ forest measurements to map the distribution of matter \cite{Hz2,Hz3}. The 31 $H(z)$ using the BAO technique covering the redshift range $0.24 \leq z \leq2.40$ are summarized
by Mukherjee and Banerjee \cite{HzBAO}. However, there may be correlations among some of the data
points in 31 BAO samples because they either belong to the same analysis or overlap. Thus, the covariances among the BAO $H(z)$ data should be taken into our analysis. The covariances are publicly available in refs \cite{HzBAO82,HzBAO84,HzBAO85,HzBAO88}. It also should be noted that the BAO data are not completely cosmological model-independent, because one needs to assume a prior on the radius of the sound horizon, which is inferred from the CMB observations. However, the prior of radius of the sound horizon obtained from the CMB analysis is only used to calibrate the BAO distance. In turn, it can be used to extract the model-dependent $H_0$ estimate, which is the basis of the so-called inverse distance ladder, which again results in a small value for the Hubble parameter, very close to the Planck value within $\Lambda$CDM model. In the work \cite{2023MNRAS.tmp.1590F} declared that this method only allows for a model-dependent determination of $H_0$, even when no specific cosmological model is assumed at late times by using, for example, cosmography. In other words, the measurement of BAO relies very weakly on the cosmological model, and only strongly on the Hubble constant $H_0$. The aim of this work is to measure  both cosmic curvature and opacity independently of the cosmological model, rather than the Hubble constant. Considering that BAO data are much more accurate and reliable, we also consider BAO measurements here for comparative analysis with CC data. The application of BAO data to the analysis of different cosmological models can be seen in refs \cite{2021MNRAS.505.2111L,2023MNRAS.522.6024B}.
For these reasons, we first used the observations from 32 CC alone (denotes as 32 CC), and then combined the 31 BAO $H(z)$ sample to conduct our analysis (although BAO data are cosmological model dependent in some extent). We denote these 63 Observation Hubble parameter Data (OHD) as 63 OHD.

Although the observations of HII regions and Hubble parameter redshifts cover each other well, there are very few of them meeting the same redshift at the same time. In our work, {the ANN} method  will be applied to achieve the reconstructions with the opacity-free expansion rate measurements, and then derive opacity-independent luminosity distances. The ANN method is a non-parameterized approach and  does not assume random variables that satisfy the Gaussian distribution,
which is a completely data driven approach. The main purpose of an ANN is to construct an approximate function or map that correlates the input vector with the output vector \cite{2015arXiv151107289C}.  The ANN is made up of neurons, which are very simple elements that receive digital input. Input and output do not need to conform to the Gaussian distribution. Furthermore, there is no need to assume a specific cosmological model. We simply train the ANN network based on the observed $H(z)$ data, and then predict the $H(z)$ at other redshifts that are not observed. In general, the artificial neural network consists of an input layer, one or more hidden layers and an output layer. In this work, the input of the neural network is the redshift $z$, while the output is the corresponding Hubble parameter $H(z)$ and its respective uncertainty $\sigma_{H(z)}$ at that redshift.  Each layer takes a vector from the previous layer as input, applies a linear transformation and a nonlinear activation function to the input, and propagates the current result to the next layer. Here, we adopt the Exponential Linear Unit (ELU) acting as activation function, which is $f(x)=\alpha (\exp(x)-1)$ with $x\leq0$, and $f(x)=x$ with $x\geq0$, where $\alpha$ is the hyper-parameter that controls the value to which an ELU saturates for negative net inputs.
The goal of ANN is to make its predicted result $\hat{H}$ to be as close as possible to the true value $H$, known as the mean absolute error loss function $\mathcal{L}$. The method used is gradient descent, that is, by constantly moving the loss value to the opposite direction of the current corresponding gradient to reduce the loss value \cite{2014arXiv1412.6980K}. The network is trained after $10^5$
iterations, to assure that the loss function no longer decreases. 
The ANN method has been gradually applied to many research fields in astronomy, and has shown excellent potential for constraining cosmological parameters \cite{ANN1,ANN2,ANN5,ANN6,ANN7}. We refer the reader to \cite{ANN6} for more details about ANN reconstructing $H(z)$ data. The work \cite{ANN7} released ANN code, and Python module called Reconstruct Functions with ANN (ReFANN)\footnote{https://github.com/Guo-Jian-Wang/refann}. We perform the reconstruction of 32 CC and 63 OHD datasets, and the final results show in Fig. 2. It should be interesting to report that the Hubble constant value is taken to be $H_0=H(z=0)=67.35\pm16.5 \Mpc$ by using ANN reconstruction with 32 CC $H(z)$ observational dataset, and this value changes to $H(z=0)=68.27\pm5.2 \Mpc$ by using ANN reconstruction with the 63 OHD dataset. These results are consistent with the value $H_0=67.4 \pm 0.5 \Mpc$ inferred from the CMB measurement using the \emph{Planck} data within $1\sigma$ confidence level \cite{2020A&A...641A...6P}, and also
support the result $H0 = 69.8\pm1.1 \Mpc$ measured from the Tip of the Red Giant Branch \cite{2019ApJ...882...34F}. Similar results are reported in many work \cite{2023PhRvD.107b3520R,2014MNRAS.441L..11B} by using Gaussian Process (GP) with CC dataset. For instance, the work \cite{2021EPJC...81..127B} found that $H(z=0)=68.57\pm1.86 \Mpc$ from the joint analysis SN+CC datasets with the GP reconstruction, which is fully consistent with our work.
These reconstruction method provides a model-independent way for extracting cosmological information to solve some questions from the observational data, such as dark energy \cite{2017PhRvD..95b3508W},  $H_0$ tension \cite{2018JCAP...05..052N,2020PhRvD.101j3505D}, cosmic growth \cite{2013PhRvD..87b3520S,2018PhRvD..97l3501J} and so on. These reconstruction methods of data may provide a new approach for the current tension. From the reconstructed results, one can see that the data uncertainty obtained by ANN method can be compared with the real observation uncertainty, which is a conservative estimate. In our reconstructed Hubble diagrams, the mean Hubble parameter is in agreement with other approach to reconstruction such as GP \cite{2021JCAP...09..014B,2012PhRvD..85l3530S}.
Therefore, the $1\sigma$ confidence region reconstructed by ANN can
be considered as the average level of observational error.
We refer the reader to Ref.~\cite{ANN6} for further details on this issue.

\begin{figure}
\begin{center}
\includegraphics[width=1\linewidth]{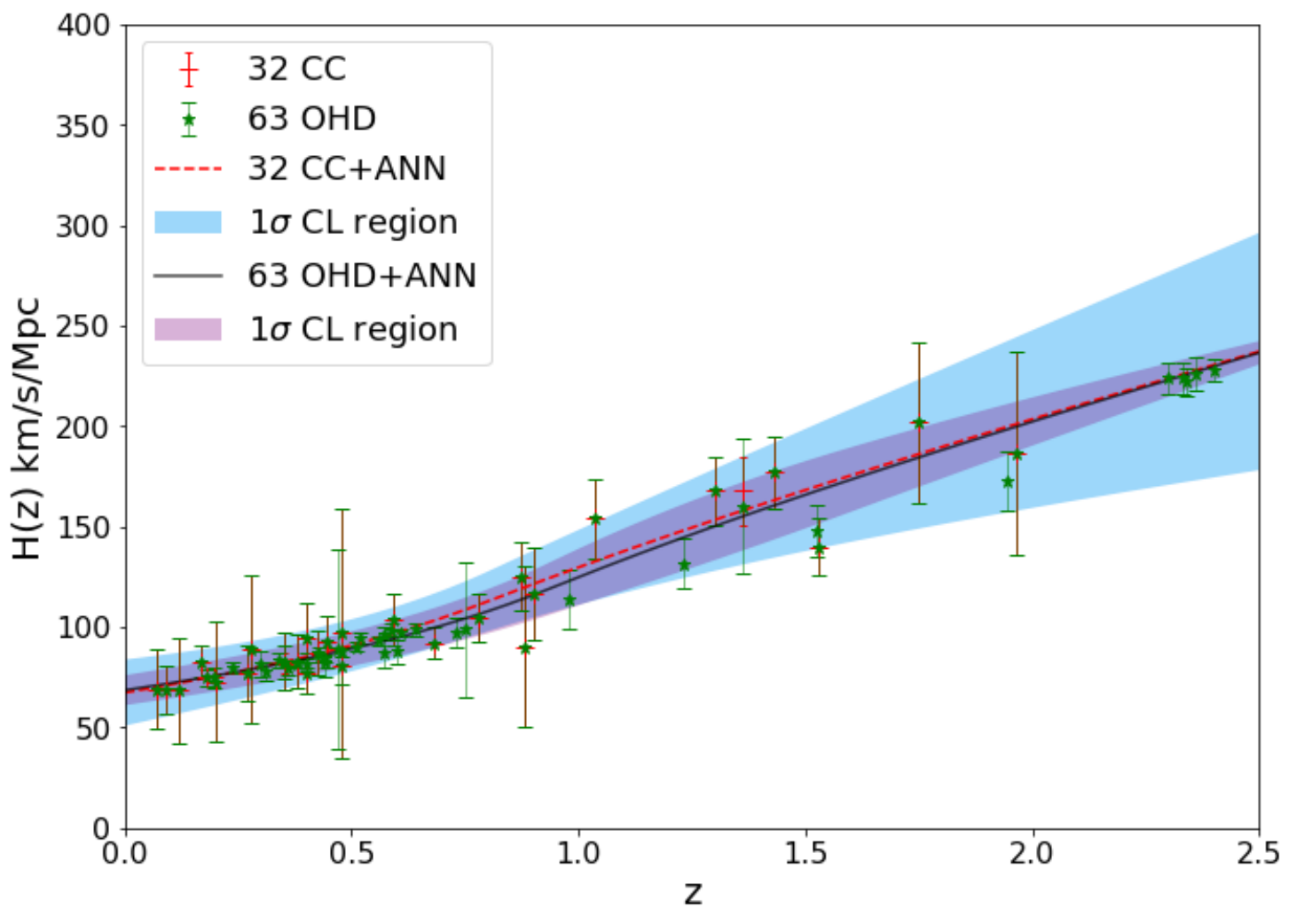}
\end{center}
\caption{{\it} The reconstructed function $H(z)$ and their corresponding $1\sigma$ errors by using ANN method with 32 CC and 63 OHD datasets. The dots with error bars represent observational Hubble parameter measurements. }
\end{figure}

For the reconstructed $H(z)$ data processing, we obtain comoving distance by using simple trapezoidal rule method
\begin{equation}
D_C\backsimeq\sum\frac{c}{H(z_i)}\Delta z_i,
\end{equation}
where $\Delta z_i=\frac{1}{2}(z_{i+1}-z_{i-1})$ is smaller enough.  Taking into account the fact that there may be correlations between the ANN reconstructed data points, and data points may be reused when using trapezoidal integrals.  Hence the covariance between these data should be considered,  and the covariance of the $D_C$ is approximately as \cite{24,26}
\begin{equation}
{\rm Cov_{D_C}}(z_i,z_j)=\sum^i_l\sum^j_k{\rm Cov}(\frac{c}{H(z_k)},\frac{c}{H(z_l)})\Delta z_k\Delta z_l.
\end{equation}
 {The further propagate to the covariance of $D_L$ is}
\begin{equation}
{\rm Cov_{D_L}}({\rm\Omega_K};z_i,z_j)=\mathcal{C}_i\mathcal{C}_j{\rm Cov_{D_C}}(z_i,z_j),
\end{equation}
where the coefficient $\mathcal{C}_i$
in error propagation is  {as} function of ${\rm\Omega_K}$,
\begin{equation}
\mathcal{C}_i = \left\lbrace \begin{array}{lll}
\frac{D_H({1+z_i})}{\sqrt{|\rm\Omega_{\rm K}|}}\coth\left[\sqrt{|\rm\Omega_{\rm K}|}{D_{C,i}/D_H}\right]~~{\rm for}~~\rm\Omega_{K}>0,\\
{D_{C,i}}{(1+z_i)}~~~~~~~~~~~~~~~~~~~~~~~~~~~~~{\rm for}~~\rm\Omega_{K}=0, \\
\frac{D_H({1+z_i})}{\sqrt{|\rm\Omega_{\rm K}|}}\cos\left[\sqrt{|\rm\Omega_{\rm K}|}{D_{C,i}}/D_H\right]~~~~{\rm for}~~\rm\Omega_{K}<0.\\
\end{array} \right.
\end{equation}

The opacity-independent luminosity $D_L$ is inferred from Eqs. (2) and (3) based on the reconstructed $H(z)$ observation. It should be noted that, instead of taking a prior value for the curvature of the universe, we fit it as a free parameter along with the opacity parameter. Thus, the luminosity distance obtained from reconstructed $H(z)$ observation includes an unknown curvature parameter.

\subsection{Simultaneous measurements on the cosmic curvature and opacity}

From the theoretical perspective, we obtain the opacity-dependent luminosity distance
$D_{L,HII}$  {via} the "$L$--$\sigma$" of HII galaxies and
extragalactic HII regions, and the opacity-independent luminosity distance $D_{L,Hz}$ can be
derived from Hubble parameter observations. In order to simultaneously measure the cosmic curvature and opacity, for a given $D_{L,HII}$ data point, the $D_{L,Hz}$ should be observed at the same redshift. If one only considers the actual observational sample, it is difficult to achieve rigorous measurements and get convincing results. In fact, it is important to note that we can reconstruct the Hubble parameters at any redshift interval and match them with the HII data. However, the redshift distribution of $H(z)$ is actually different from the redshift distribution of HII region, and there is no reason to choose the observed redshift of HII region to match the reconstructed $H(z)$ data. To avoid introducing additional systematic errors, a cosmological model-independent selection criterion is considered. We take $|z_{HII}-z_{Hz}|<0.005$ in our analysis \cite{6,7}.

Once we have two sets of luminosity distances at the same redshift, one of them contains cosmic opacity from observations in the HII region, {and the other one} that does not contain cosmic opacity but contains cosmic curvature from reconstructed $H(z)$ observation. We can directly perform a model-independent measure for cosmic opacity parameter $\tau$ and curvature $\rm\Omega_K$, which is given by the following form \cite{53,54}
\begin{equation}\label{eq:DDR}
D_{\rm L,\rm Hz}(z)=D_{\rm L,\rm HII}(z)e^{\tau/2},
\end{equation}
where an opacity parameter $\tau$ {is} introduced to describe the optical depth associated with cosmic absorption quantifying how opaque the universe is. Note that any statistically significant deviation from $\tau=0$ could indicate an opaque universe assumption.
For the optical depth, we use $\epsilon$ to characterize the cosmic opacity, and note that for small $\epsilon$ {which} is equivalent to {assume} an optical depth parameterization $\tau(z)=2\epsilon z$. {In Ref.}\cite{2009JCAP...06..012A}, authors have verified that using $\epsilon$ instead of $\tau$ has {considerable} accuracy. In the {investigations} on the opacity of the late universe, the constraint on cosmic opacity $\epsilon$ indicated that it is a small quantity by using such parameterization form \cite{2010JCAP...10..024A,Li13,2015PhRvD..92l3539L}.

Then we can maximize the described logarithmic likelihood function by performing Markov chain Monte Carlo (MCMC) implementations. The log-likelihood function is given by
\begin{equation}
\ln \mathcal{L}=-\frac{1}{2}\sum^N2\pi\ln (\det\mathbf{Cov})-\frac{1}{2}\rm{\Delta} \mathbf{D}_L^T \mathbf{Cov}^{-1}\Delta \mathbf{D_L},
\end{equation}
where $\mathbf{D_L}$ is the luminosity distance difference vector
\begin{equation}
\rm{\Delta } \mathbf{D_L}=\mathbf{D}_{\rm{L},\rm{HII}}(\epsilon;\mathbf{z}_{\rm{HII}})-
\mathbf{D}_{\rm{L},\rm{Hz}}^{rec}(\rm\Omega_K;\mathbf{z}_{\rm{Hz}}),
\end{equation}
where $\mathbf{Cov}$ is the corresponding total covariance matrix
\begin{equation}
\mathbf{Cov}=\mathbf{Cov}^{rec}_{D_{\rm{L,Hz}}}+ \sigma^2_{D_{\rm{L,HII}}} \mathbf{I},
\end{equation}
where the $\mathbf{I}$ is unit matrix, the $\mathbf{Cov}^{rec}_{D_{\rm{L,Hz}}}$
is covariance matrix of the reconstructed luminosity distance from $H(z)$ data in Eq (9), and the $\sigma^2_{D_{\rm{L,HII}}}$ is the uncertainty of the HII regions, including velocity dispersion, flux density observational uncertainty, intercept uncertainty and slope uncertainty, which are calculated by the standard error transfer formula. We use the Python module $emcee$ \cite{55} to perform the MCMC analysis.

\section{Results and Discussion}

\begin{figure*}
\begin{center}
{\includegraphics[width=0.32\linewidth]{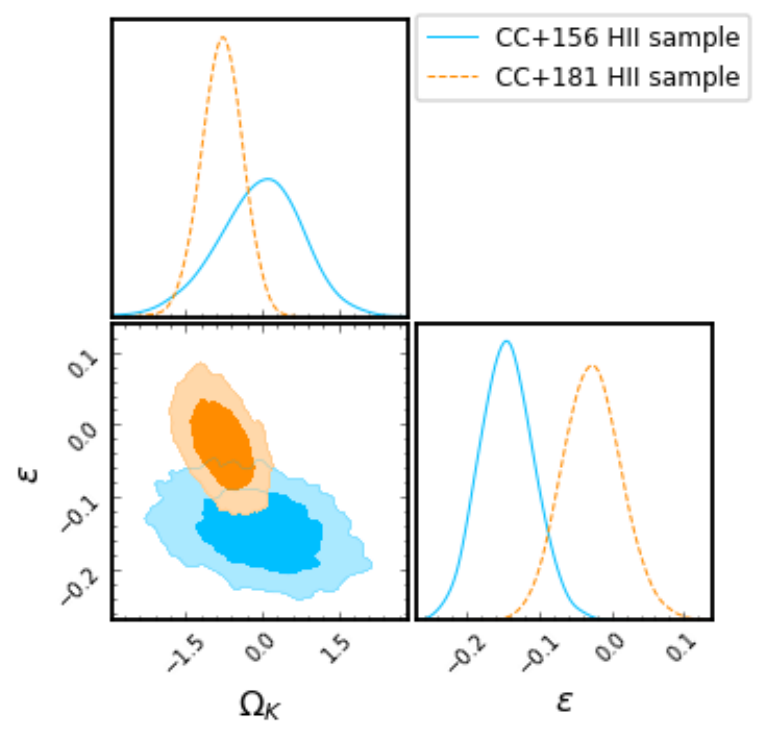}
\includegraphics[width=0.32\linewidth]{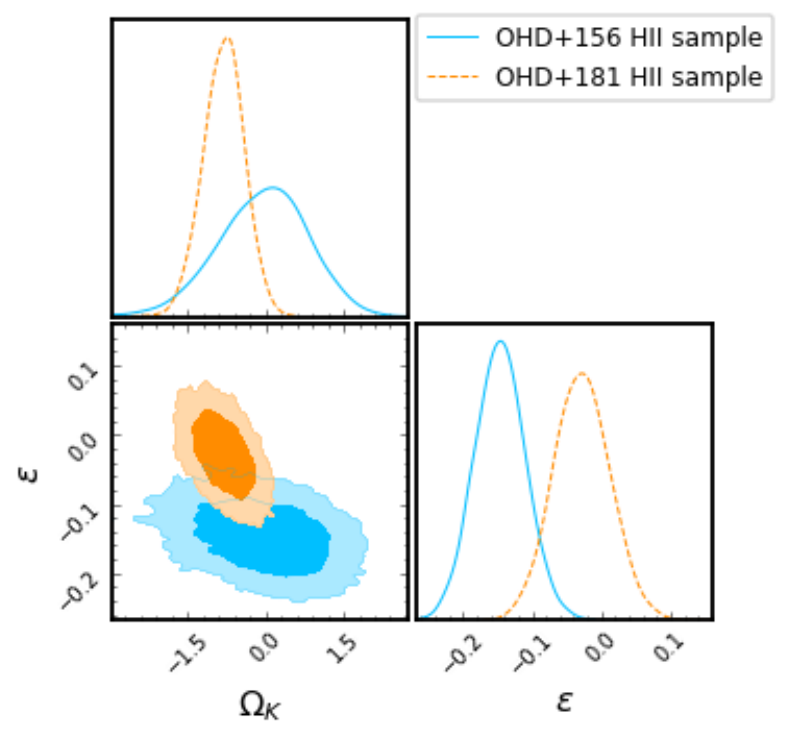}
\includegraphics[width=0.32\linewidth]{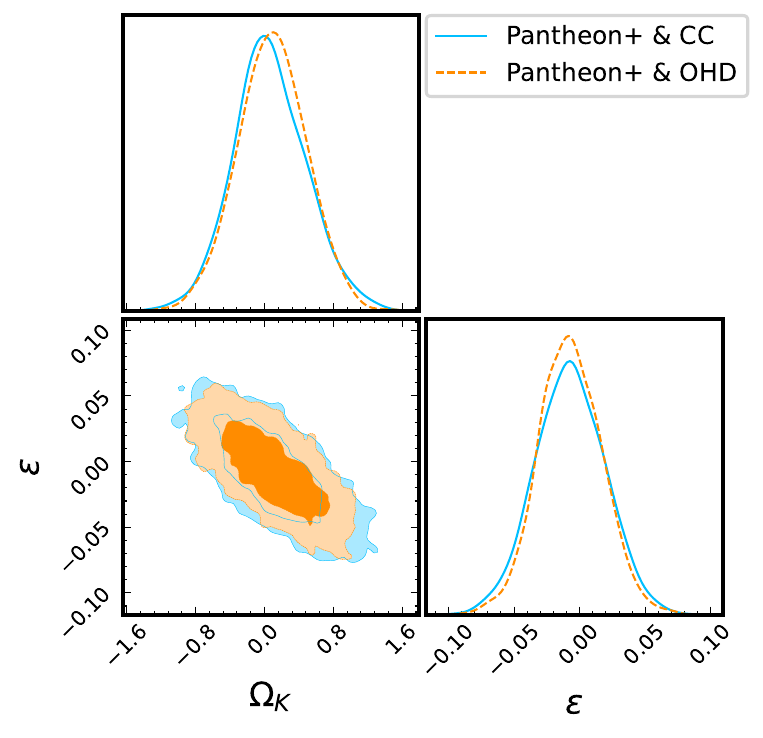}}
\end{center}
\caption{ The 1D and 2D marginalized probability distributions for
the cosmic opacity parameter $\epsilon$ and curvature parameter $\rm\Omega_K$ by using ANN
reconstructed different observation datasets.}
\end{figure*}

\begin{table}
\renewcommand\arraystretch{1.8}
\caption{\label{tab:result} Summary of the constraints on the spatial curvature parameter $\rm\Omega_K$ and cosmic opacity parameter $\epsilon$ by using different data combination.}

\begin{center}
\begin{tabular}{l| c| c }
\hline
\hline
Data Combination  & $\rm\Omega_K$ &$\epsilon$  
\\
\hline
CC+156 HII  & $0.016^{+0.762}_{-0.884}$ & $-0.145^{+0.036}_{-0.036}$ \\
\hline
CC+181 HII  & $-0.800^{+0.384}_{-0.389}$ & $-0.028^{+0.040}_{-0.040}$ \\
\hline
OHD+156 HII  & $-0.007^{+0.793}_{-0.845}$ & $-0.146^{+0.036}_{-0.035}$ \\
\hline
OHD+181 HII  & $-0.811^{+0.389}_{-0.391}$ & $-0.029^{+0.040}_{-0.038}$ \\
\hline
CC+SN Ia  & $0.043^{+0.446}_{-0.380}$ & $-0.007^{+0.026}_{-0.026}$ \\
\hline
OHD+SN Ia  & $0.094^{+0.395}_{-0.393}$ & $-0.008^{+0.024}_{-0.022}$  \\
\hline
\hline
\end{tabular}
\end{center}
\end{table}

Let's start with the ANN reconstructed 32 CC $H(z)$ observational data. It shoud be stressed that the $H_0$ value is adopted from the the ANN reconstruction at $z=0$ with the 32 CC $H(z)$ observation in this framework, i.e., $H_0=67.35\pm16.5 \Mpc$.  Combined with 156 HII regions sample, the 1D marginalized probability distributions and 2D regions with $1\sigma$ and $2\sigma$ contours corresponding to cosmic opacity and curvature parameters are shown in Fig.~3. The best-fitting values with the corresponding $1\sigma$ uncertainties are $\rm\Omega_K =0.016^{+0.762}_{-0.884}$  and $\epsilon=-0.145^{+0.036}_{-0.036}$. Meanwhile, for the latest 181 HII regions sample, we obtain the
best-fitting values $\rm\Omega_K$ and $\epsilon$ with 1$\sigma$ confidence level are  $\rm\Omega_K =-0.800^{+0.384}_{-0.389}$  and $\epsilon=-0.028^{+0.040}_{-0.040}$.  In Fig.~4, the $1\sigma$ and $2\sigma$ confidence level contours for parameter estimations represent the constraint results. The  numerical results of the constraints on $\rm\Omega_K$ and $\epsilon$ are shown in Table 1.  The best-fitting values for both the cosmic curvature and opacity are negative, {and} our results support a transparent and closed universe at $1\sigma$ confidence level.

Working on the ANN reconstructed 63 OHD $H(z)$ sample, the Hubble constant value is taken to be $H_0=68.27\pm5.2 \Mpc$ from the ANN reconstruction at $z=0$, and we adopt this value for a prior. In this case,
the result seems to support a slight preference for the nonzero cosmic opacity ($\epsilon=-0.146^{+0.036}_{-0.035}$) within 1$\sigma$ confidence level by combining 151 HII regions sample, while supporting a flat universe in this case. However, combining 181 HII  regions sample, we obtain a closed universe ($\rm\Omega_K=-0.811^{+0.389}_{-0.391}$) within $1\sigma$ confidence level.  In this case, the result indicates that a transparent universe is supported.  The corresponding numerical results are displayed in lines 3 and 4 of Table 1.  Recent reports suggest that there may be inconsistencies in the spatial curvature between the early and late universe, i.e., combining the Planck temperature and polarization power spectra data, the work showed that a closed universe ($\rm\Omega_K=-0.044^{+0.018}_{-0.015}$) was supported. However, with the combination of the Planck lensing data and low redshift baryon acoustic oscillations observation, a flat universe was precisely constrained within $\rm\Omega_K=0.0007\pm0.0019$  \cite{2021APh...13102605D,2020NatAs...4..196D,2021PhRvD.103d1301H}.

Since the large of deviations of the results with the recent work \cite{7,Liao13,LiuT21b}, this prompted one thought about the above results: whether the HII region samples are robust and reliable enough to support us to infer phenomenology?
In the subsequent analysis, we combine the ANN reconstructed $H(z)$ data and latest SN Ia observational sample released by the SH0ES and Pantheon$+$ collaboration to constrain on cosmic opacity and curvature parameters \cite{2022ApJ...938..113S}. The Pantheon$+$ dataset includes 1701 light curves, spectroscopically confirmed SN Ia and covering the redshift range $0.001<z< 2.26$ which comes from 18 different surveys. For SN Ia dataset, the value of absolute magnitude $M_B$ is important. The $M_B$ is usually calibrated by local measurement such as Cepheid variable stars. However, there is strong degeneracy  between the $M_B$ and $5\log_{10}H_0$ term inside the distance modulus. Considering the consistency of the work, we convert the Hubble constant used to absolute magnitudes, i.e., $M_B=-19.435$ $mag$ corresponds to $H_0=67.35 \Mpc$ with the ANN reconstructed 32 CC dataset, and $M_B=-19.406$ $mag$ corresponds to $H_0=68.27 \Mpc$ with the ANN reconstructed 63 OHD dataset.
The $1\sigma$ and $2\sigma$ confidence level contours for $H(z)+$ SN Ia constraint is shown in the right panel of the Fig. 3.  We obtain the best-fitting values are $\rm\Omega_K=0.043^{+0.446}_{-0.380}$ and $\epsilon=-0.007^{+0.026}_{-0.026}$ by using the ANN reconstructed CC and the Pantheon$+$ datasets. Working on 63 OHD sample, the best-fitting values are $\rm\Omega_K=0.094^{+0.395}_{-0.393}$ and $\epsilon=-0.008^{+0.024}_{-0.022}$.
This result suggests that, combined with the most popular astronomical observations of SN Ia plus $H(z)$, our universe is a flat and transparent Universe. Compared with results from the HII regions datasets, we find that
the deviation of the cosmic curvature and opacity parameters from zero values is largely due to the HII samples. Although the samples of HII region are not sufficiently reliable (it may be that the small sample size and large uncertainty lead to the bias of the inferred phenomenology), this improved model-independent approach proposed in this work to infer phenomenology may have a place in the future.

The second question is whether the degeneracy between curvature and opacity reinforces the negative curvature results obtained from the HII region datasets.
In order to better analyze the degeneracy between curvature and opacity, we first assume a flat universe ($\rm\Omega_K=0$). Combining the results given by CC+156 HII and OHD+156 HII datasets, we are able to conclude that the universe is opaque at $1\sigma$ confidence level, and the numerical results are given in Table 2. The results given by CC+181 HII and OHD+181 HII are shown a more opaque universe at $1\sigma$ confidence level compared to the case of freeing the cosmic curvature parameter. We plot the probability density function of the cosmic opacity parameter in Fig. 5. This result is perfectly consistent with our hypothesis. When true HII galaxies emit photons travel through the universe to the observer, the flux received by the observer decreases because the universe may be opaque, which leads to an increase in the actual observed luminosity distance.  However, the observations of SN Ia plus $H(z)$ give a almost transparency Universe. This results is in perfect agreement with the recent work \cite{7,LiuT21b,2015PhRvD..92l3539L}. Thus, we are able to conclude that this phenomenological deviation ($\epsilon\neq0$) is mainly due to the unreliability or bias of the HII region datasets, rather than the CC and OHD samples.

Assuming a transparent universe ($\epsilon=0$), we also plot the probability density as the function of the cosmic curvature parameter in  Fig. 6, and the numerical results are shown in Table 2. {We find that if we increase the degree of freedom of the opacity parameter of the universe, the curvature of the universe has a larger negative value.} From above results, there is little difference between the OHD and CC samples, however, the difference between the 156 HII and 181 HII samples is very large, which will lead to different cosmic opacity and curvature.
Obviously, the difference between the two HII samples has a significant effect on the curvature constraint, and the curvature value of the sample in the 181 HII region is greater than the negative curvature value of the sample in the 156 HII region. Working on the Pantheon+ with CC and OHD datasets, a flat Universe is both supported within the 1$\sigma$ uncertainties, but the center value is a little negative. This result is consistent with the  value inferred from Planck data \cite{2020A&A...641A...6P}. These results strengthen our argument in answer to the first question that the HII region samples are not particularly reliable.
However, although the degeneracy between the cosmic opacity and the spatial curvature still exists, our method  provides a new approach for constraining both the cosmic opacity and the spatial curvature, and alleviates this situation to some extent compared to considering cosmic curvature or opacity alone.

\begin{figure}
\begin{center}
\includegraphics[width=0.85\linewidth]{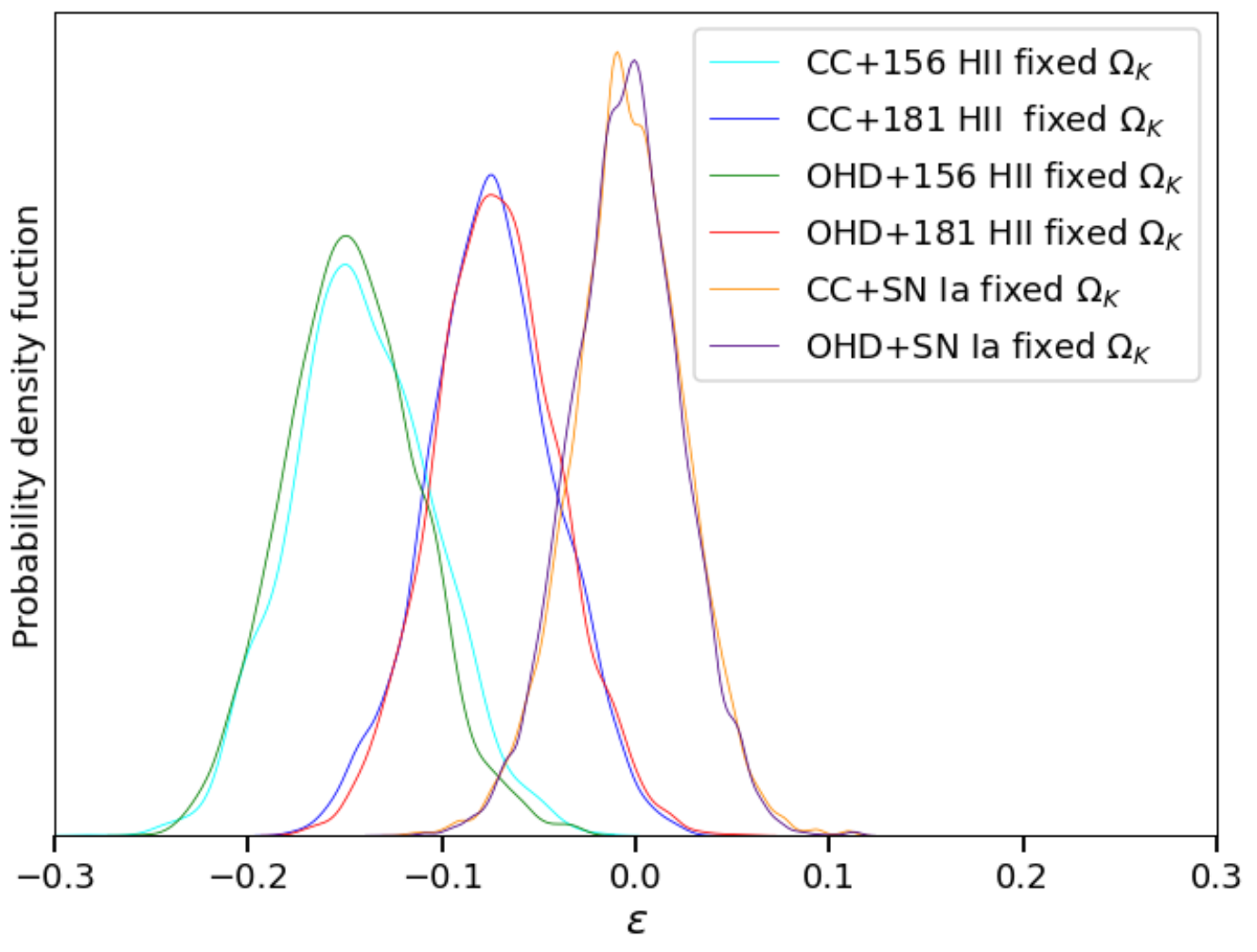}
\end{center}
\caption{The probability density function of the cosmic opacity parameter with a flat universe ($\rm \Omega_K=0$).}
\end{figure}

\begin{figure}
\begin{center}
\includegraphics[width=0.85\linewidth]{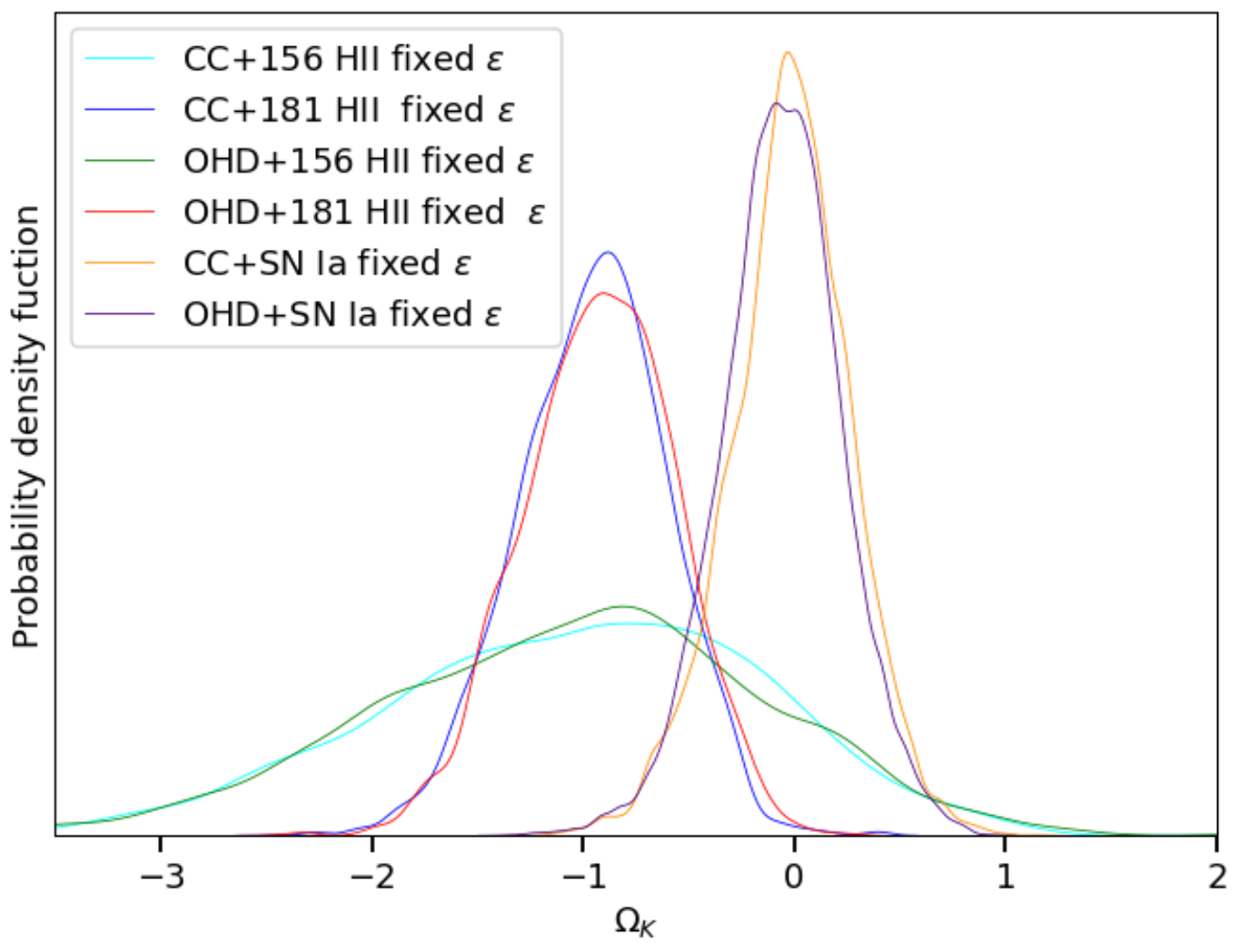}
\end{center}
\caption{The probability density function of the cosmic curvature with a transparent universe ($\epsilon=0$).}
\end{figure}
\begin{table}
\renewcommand\arraystretch{1.5}
\caption{ Summary of the constraints on the spatial curvature parameter $\rm\Omega_K$ and cosmic opacity parameter $\epsilon$ by using different data combination, with the assumption of  a flat universe ($\rm \Omega_K=0$) or a transparent  universe ($\epsilon=0$).}

\begin{center}
\begin{tabular}{l| c| c }
\hline
\hline
Data Combination  & $\rm\Omega_K$ &$\epsilon$  
\\
\hline
CC+156 HII (fixed $\rm\Omega_K$) & $-$ & $-0.143^{+ 0.039}_{-0.034}$ \\
\hline
CC+156 HII (fixed $\rm\epsilon$) & $ -0.988^{+0.847}_{-0.985}$ & $-$ \\
\hline
CC+181 HII (fixed $\rm\Omega_K$) & $-$ & $-0.075^{+0.034}_{-0.032}$ \\
\hline
CC+181 HII (fixed $\rm\epsilon$) & $-0.937^{+0.332}_{-0.368}$ & $-$ \\
\hline
OHD+156 HII (fixed $\rm\Omega_K$) & $-$ & $-0.148^{+0.036}_{-0.033}$ \\
\hline
OHD+156 HII (fixed $\rm\epsilon$) & $-1.003^{+0.857}_{-0.995}$ & $-$ \\
\hline
OHD+181 HII (fixed $\rm\Omega_K$) & $-$ & $-0.072^{+ 0.033}_{-0.031}$ \\
\hline
OHD+181 HII (fixed $\rm\epsilon$) & $-0.908^{+ 0.342}_{-0.382}$ & $-$ \\
\hline
CC+SN Ia  (fixed $\rm\Omega_K$)& $-0.012^{+0.269}_{-0.292}$ & $-$ \\
\hline
CC+SN Ia (fixed $\rm\epsilon$)& $-$ & $-0.003^{+0.019}_{-0.018}$ \\
\hline
OHD+SN Ia (fixed $\rm\Omega_K$)& $-0.056^{+0.266}_{-0.280}$ & $-$ \\
\hline
OHD+SN Ia (fixed $\rm\epsilon$)& $-$ & $-0.003^{+0.018}_{-0.019}$ \\
\hline
\hline
\end{tabular}
\end{center}
\end{table}
Benefiting from the ANN reconstructed technology, the $\rm {Hz/HII}$ data pairs satisfying the redshift selection criteria have a considerable growth. Therefore, a considerable amount of high-redshift
samples (beyond the redshift limit of SN Ia $z>1.4$) have been included in our analysis. Actually, the combination of Hubble parameter and the HII regions observation allows us to use ANN reconstructed technology to get more precise measurements on the cosmic opacity parameter at level $\rm\Delta\epsilon\sim 10^{-2}$, although the measurement precision of the curvature is at the level of $\rm\Delta\Omega_K\sim 10^{-1}$.
Our method {provides} a model-independent constraint both on cosmic opacity and curvature, more stringent than other current results based on real observational data. In addition, our results show that there is a very strong degeneracy between cosmic opacity and curvature, and this degeneracy is negatively correlated.

In order to highlight the potential of our method, it is necessary
to compare our results {with previous works}.  Many works turned to luminous sources with known (or standardizable) intrinsic luminosity in the universe, like SN Ia, quasars and so on, to infer observations of opacity-dependent luminosity distances. For example, some works \cite{Li13,Liao13,2015PhRvD..92l3539L} used Union 2.1 + galaxies cluster sample, Union2.1 + $H(z)$, JLA + $H(z)$ to constrain cosmic opacity, and obtained the precision on cosmic opacity parameter at level $\rm\Delta\epsilon\sim 10^{-1}$. The recent works \cite{54,fu21,geng,wei1} used simulated gravitational wave observations to replace Hubble parameter observations to constrain cosmic opacity. {These} works showed the opacity parameter $\Delta \epsilon \sim 10^{-3}$ at 68.3\% confidence level.  However, we should seek other methods and technologies
until the observed gravitational wave events will be sufficient to get statistical results in the
future.

\section{Conclusion}

The cosmic opacity and curvature parameter both play the important roles in modern cosmology. The cosmic opacity may indicate that photon numbers are not conserved from source to the observer due to some new physical phenomenon. Cosmic opacity may be caused by absorption/scattering due
to matter in the universe, or by extragalactic magnetic fields that can turn photons into unobserved
particles (e.g. light axions, chameleons, gravitons, Kaluza-Klein modes), and it thus is crucial to correctly interpret cosmic opacity for astronomical photometric measurements, like SN Ia, HII regions and quasar observations. According to general relativity, photons move along geodesics in the universe, and curvature determines the spatial structure of the universe, which means that the path of photons from the source to the observer will be affected by the curvature of the universe. Different curvature of the universe will result in different measured cosmological distances, which may also cause the supernova to faint.  Therefore, it is necessary to measure cosmic curvature and opacity simultaneously.

In this paper, we have proposed a new model-independent method to simultaneously measure the cosmic curvature and opacity by using the latest observations of HII galaxies acting as standard
candles and the latest Hubble parameter observations.  We adopt the non-parameterized method Artificial Neural Network to reconstruct observed Hubble parameter. Our results support a slightly opaque and closed universe at $1\sigma$ confidence level.
However, it should be emphasized that the cosmic absorption (caused by the  opaque of the Universe) affects the luminosity distances derived from ``$L$--$\sigma$" relation of HII regions, but also generates
influences on the measured luminosity distances of SN Ia. In particular, considering the fact that a opaque Universe will lead an increase in the actual observed luminosity distance, which means that the sources we observe are farther away and less bright than they really are. Therefore, this is also a mechanism to explain the dimming of distant SN Ia.
In the subsequent analysis, we suggest that there could be two possible reasons for this results. Firstly, the increase in the degree of freedom of the opacity parameter makes curvature more negative, that is, the negative correlation between the opacity and the curvature parameter leads to this result. Secondly, unreliability of the two HII region samples rather than the $H(z)$ samples, cosmic curvature obtained using the HII sample has the large negative values. This conclusion is supported by jointly analysis of the most popular astronomical observations of SN Ia plus $H(z)$ dataset, such combination supports a flat and transparent Universe within 1$\sigma$ uncertainty.
More importantly, we obtain the measurement precision on the cosmic opacity parameter $\rm\Delta\epsilon\sim 10^{-2}$, and the measurement precision on curvature $\rm\Delta\Omega_K\sim 10^{-1}$, simultaneously. In the case of assuming a transparent universe, our results indicate in some ways that our universe has a spatial structure of negative curvature.  A strong degeneracy between the cosmic opacity and curvature parameters is also revealed in this analysis. Note that such negative correlation could potentially affect our constraint on cosmic opacity or curvature alone. Meanwhile, we need to emphasize that our work suggests a new approach to constrain the cosmic opacity and the spatial curvature, simultaneously. Although the degeneracy between the cosmic opacity and the spatial curvature still exists, our method alleviates this situation to some extent compared to considering cosmic curvature or opacity alone.

The ANN method we used here is not unique. It is still worth exploring whether these conclusions will change with different machine learning reconstruction methods.
However, both GP and ANN method have their own advantages and disadvantages \cite{ANN7}, and show great potential in the studies of precision cosmology. For instance, the GP reconstruction greatly reduces the uncertainty of data, and the reconstructed data points are related to each other. The ANN method is more like a black box, and we don't know what's going on inside the box. Therefore, we emphasize here that in the era before the wealth of available data, it is also very necessary to carefully choose the method of data reconstruction.

As a final remark,  although the simultaneous constraint on cosmic opacity and curvature by using HII regions and Hubble parameter measurements do not significantly improve precision,  yet
it helps us to gain a deeper understanding relation between cosmic opacity and  curvature in
the early universe ($z\sim2.5$). We also look forward to a large amount of future data, not only from the Hubble parameters, but also from the HII regions, allowing us to further improve the precision of the cosmic opacity and curvature constraints through various machine learning techniques in the future.

\section*{Acknowledgments}
The authors are grateful to the referee for constructive comments, which allowed to improve the paper substantially.
Liu. T.-H was supported by National Natural Science Foundation of China under Grant No. 12203009;  Chutian Scholars Program in Hubei Province (X2023007); Hubei Province Foreign Expert Project (2023DJC040). Y. Y. was supported by the National Natural Science
Foundation of China under Grant No. 12105097, and Scientific Research Fund of Hunan Provincial Education Department(No. 20C0787). Wu. S.-M was supported by National Natural Science Foundation of China under Grant No. 12205133.

\bibliography{references}

\end{document}